\documentclass[aps,pra,english,twocolumn,showpacs,preprintnumbers,amsmath,amssymb,floatfix]{revtex4}
\usepackage[T1]{fontenc}
\usepackage[latin9]{inputenc}
\usepackage{graphicx}
\usepackage{amssymb}
\usepackage{babel}


\usepackage{babel}
\makeatother

\begin{document}

\title{Casimir experiments showing saturation effects}

\author{Bo E. Sernelius}

\affiliation{Division of Theory and Modeling, Department of Physics, Chemistry
and Biology, Link\"{o}ping University, SE-581 83 Link\"{o}ping, Sweden}

\email{bos@ifm.liu.se}

\begin{abstract}
We address several different Casimir experiments where theory and
experiment disagree.  First out is the classical Casimir force measurement
between two metal half spaces; here both in the form of the torsion
pendulum experiment by Lamoreaux and in the form of the Casimir pressure
measurement between a gold sphere and a gold plate as performed by Decca et
al.; theory predicts a large negative thermal correction, absent in the
high precision experiments.  The third experiment is the measurement of
the Casimir force between a metal plate and a laser irradiated
semiconductor membrane as performed by Chen et al.; the change in force
with laser intensity is larger than predicted by theory.  The fourth
experiment is the measurement of the Casimir force between an atom and a
wall in the form of the measurement by Obrecht et al.  of the change in
oscillation frequency of a ${}^{87}Rb$ Bose-Einstein condensate trapped to
a fused silica wall; the change is smaller than predicted by theory.  We
show that saturation effects can explain the discrepancies between theory
and experiment observed in all these cases.
\end{abstract}


\maketitle

\section{Introduction}

The Casimir force in the purest geometry is extremely fascinating: There is
an attractive force between two parallel, perfectly reflecting metal plates
embedded in vacuum; this geometry was treated by Casimir in his classical
paper \cite{Casi} of 1948.  One may say that the fact that Maxwell's
equations have solutions in vacuum, that electromagnetic waves need no
medium to propagate in, gives rise to these forces.  One way to calculate
the forces is to calculate the energy in the system and then take minus the
derivative with respect to separation, $d$, between the plates.  At zero
temperature the energy may be obtained as a summation of the zero-point
energy of all the electromagnetic normal modes of the system.  The normal
modes change when $d$ changes and so do the zero-point energies.  At
finite temperature the actual population of the modes affects the forces.

If the idealized perfectly reflecting metallic plates are replaced by
plates of real materials the boundary conditions are different.  There is
still a force, but it is now different.  For real metals the vacuum modes
penetrate to some extent into the plates and furthermore new modes appear,
modes that are bound to the surfaces of the plates; these are evanescent
waves, so-called surface modes \cite{Ser}.  For large separations the
vacuum modes, or propagating waves, dominate and the force, at least at
zero temperature, is identical to the force in the idealized system.  For
smaller separations the surface modes dominate and the force acquires
different characteristics; in this range the force is called the van der
Waals force.

In a pioneering work Sparnaay \cite{Spar} tried to verify the existence of
the Casimir force experimentally between two parallel plates.  However, the
uncertainties were of the same order of magnitude as the force itself so the experiment
was non-conclusive.  The interest in the Casimir force virtually exploded
a decade ago.  This increase in interest was spurred by the torsion
pendulum experiment by Lamoreaux \cite{Lamo}, which produced results with
good enough accuracy for the comparison between theory and experiment to be
feasible.  This stimulated both theorists \cite{BosSer, LamRey, Bordag,
BreAar, Svet, Nogu} and experimentalists \cite{Mohi, Ede, DecLop, Bres} and
the field has grown constantly since then.  Another reason for this
increase of interest in Casimir forces is the huge shift of general focus
of the science community into nanoscience and nanotechnology where these
forces become very important.  

A setback came immediately.  Theory and experiment agree quite well for low
temperatures, but at room temperature, where most experiments are performed
there are serious deviations.  Each new type of experiment has lead to new
puzzling discrepancies between theory and experiment.  In the torsion
pendulum experiment by Lamoreaux \cite{Lamo} and in the micromachined
torsion oscillator experiment by Decca et al.  \cite{Decca} the Casimir
force and Casimir pressure, respectively, were found to agree much better
with the theoretical zero temperature results than with the room temperature
results although both experiments were performed at room temperature.  The
problems appeared when one improved the treatment of the dielectric
properties of the metals by inclusion of dissipation \cite{BosSer}; this
can be done either by using tabulated, experimental optical data or by using
the Drude dielectric function.  The origin of the dissipation is carrier
scattering against impurities and lattice imperfections; at finite
temperature also carrier-phonon scattering contributes.  Removing
dissipation from the Drude dielectric function, i.e., using the so-called
Plasma model has the effect that the discrepancies go away.  One was
facing the dilemma that improving the theory led to worse results.  There
were also claims that the theory did not obey the Nernst heat theorem, the
second law of thermodynamics.  This could, however, be shown not to be the
case \cite{SerBos, BosSer2, Brev}.  There still remained a problem for
perfect crystals where the dissipation comes from phonon scattering, only. 
This was, however resolved by inclusion of spatial dispersion \cite{Ser2};
it turned out that spatial dispersion had a very small effect on the
Casimir force, but it removed the remaining problem with the Nernst heat
theorem for perfect crystals.

The sphere and plate geometry used in these two experiments has the benefit 
compared to the two plate geometry used in Refs. \cite{Spar, Bres} that 
one avoids the alignment problem. However, one has to rely on the so-called 
proximity force approximation (PFA) \cite{Derj} in the interpretation of the 
experiments and that could be a source of error. Fortunately, the PFA, 
although based on a rather shaky foundation, seems to work quite well \cite{SerRom}.

To get a better insight into what goes wrong, or where, it would be very
useful to have experiments performed in the whole temperature range from
very low temperature up to room temperature or even higher, keeping
everything else fixed.  This is very difficult to do in practice.  Chen et
al.  \cite{Chen} came up with a very clever idea about how to collect
complementing information.  By using a gold sphere and a laser irradiated
semiconductor wafer one could study how the force changed with carrier
density in the wafer, keeping everything else fixed.  

Here one was facing
another setback. The change in force in the presence of light compared to 
in the absence of light was larger than expected. In the previous 
experiments one could get agreement between experiment and theory by 
neglecting the dissipation. In this experiment it did not matter if 
dissipation was included or not. One could get good agreement if one 
completely neglected the very few carriers, present in the absence of 
light but kept the contribution from the laser excited carriers in the 
presence of light. Thus, it seemed to be quite different reasons for the 
discrepancies in the two types of experiment.

Another experiment \cite{Ante} where the Casimir force can be measured is based on
Bose-Einstein condensates (BEC) near a surface.  For a harmonically trapped
BEC placed at a distance $d$ from the surface of a substrate the centre-of-mass 
oscillation frequency changes due to the surface-atom potential. 
Experiments were performed \cite{Obrecht} on a ${}^{87}Rb$ BEC trapped to
a fused silica wall. Theory predicted a too big a change in frequency. Good 
agreement was obtained if one neglected the contribution to the dielectric 
properties of the fused silica from the very small amount of thermally 
excited carriers in the material \cite{KliMos}. 

Thus, in the last two experimental setups it seems as if the dielectric
properties of the carriers in a semiconducting system with very small
carrier concentration have to be modified.  This has been done in two
recent publications \cite{Pita, DalLam} but if this resolves the problems
is not so clear. 

In this work we put forward what we think is the solution to all the
problems discussed above, or at least the first step towards a solution.  It
is applicable to all the discussed experimental setups.  The idea is that 
the discrepancies are due to saturation effects. We have previously
discussed these ideas in a very short and condensed article \cite{Ser4}. 
Here, we make a much more detailed presentation. In Sec. \ref{basics} we 
present the formalism that is common to all three experimental geometries. 
Then all three geometries are treated one by one, starting by two metal 
plates in Sec. \ref{G1}, followed by the semiconductor and metal 
plates in Sec. \ref{G2}, and ending by the atom wall geometry in Sec. 
\ref{G3}. Finally, we end with summary and discussions in Sec. \ref{summary}.

\section{\label{basics}Basic Formalism}

\begin{figure}
\includegraphics[width=6cm]{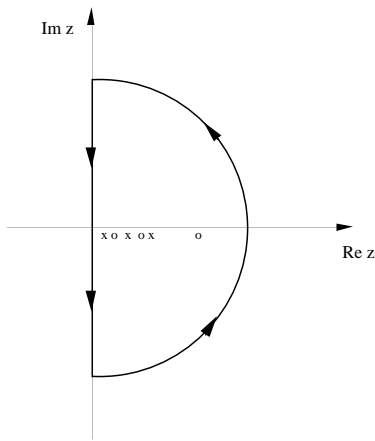}
\caption{Integration contour in the complex $z$-plane suited for zero 
temperature calculations.  Crosses and circles
are poles and zeros, respectively, of the function $f(z)$.  The radius of the
circle is let to go to infinity.}
\label{figu1}
\end{figure}

\begin{figure}
\includegraphics[width=6cm]{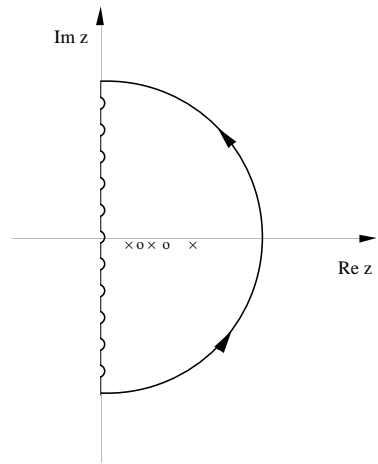}
\caption{Integration contour in the complex $z$-plane suited for finite
temperature calculations.  Crosses and circles are poles and zeros,
respectively, of the function $f(z)$.  The small semi circles are centered
at the poles of the coth function in the integrand. The radius of the
large semi circle is let to go to infinity.}
\label{figu2}
\end{figure}

The formalism we use is based on the electromagnetic normal modes of the
system \cite{Ser}.  At zero temperature the force, $F(d)$, between two objects as a
function of separation, $d$, is given by the derivative of the energy, $E$:
\begin{equation}
F\left( d \right) =  - \frac{{{\rm{d}}E\left( d \right)}}{{{\rm{d}}d}},
\label{equ1}
\end{equation}
where the energy is the sum of the zero-point energies of the modes,
\begin{equation}
E\left( d \right) = \sum\limits_i {{\textstyle{1 \over 2}}\varepsilon _i
\left( d \right)}.
\label{equ2}
\end{equation}
In a simple system with a small number of well-defined modes this summation
can be performed directly.  In the general case this is not so.  Then we
can rely on an extension of the so-called Argument Principle 
\cite{Ser,BrevEl} to find the
results.  Let us study a region in the complex frequency plane.  We have
two functions defined in this region; one, $\varphi {\rm{(z)}}$, is
analytic in the whole region; one, $f(z)$, has poles and zeros inside the
region.  The following relation holds for an integration path around the
region:
\begin{equation}
\frac{1}{{2\pi i}}\oint {dz\varphi \left( z \right)\frac{d}{{dz}}\ln
f\left( z \right)} = \sum {\varphi \left( {z_o } \right)} - \sum {\varphi
\left( {z_\infty } \right)},
\label{equ3}
\end{equation}
where $z_0 $ and $z_\infty $ are the zeros and poles, respectively, of
function $f(z)$.  In the Argument Principle the function $\varphi
{\rm{(z)}}$ is replaced by unity and the right hand side then equals the
number of zeros minus the number of poles of the function $f(z)$ inside
the integration path.  If we choose the function $f(z)$ to be the function
in the defining equation for the normal modes of the system, $f\left(
{\omega _i } \right) = 0$, the function $\varphi \left( z \right)$ as
${{\hbar z} \mathord{\left/ {\vphantom {{\hbar z} 2}} \right. 
\kern-\nulldelimiterspace} 2}$ , and let the contour enclose all the zeros 
and poles of the function $f(z)$ then Eq.  (\ref{equ3}) produces the energy
in Eq. (\ref{equ2}).  By using this theorem we end up with integrating along
a closed contour in the complex frequency plane.  In most cases it is
fruitful to choose the contour shown in Fig.  \ref{figu1}.
We have the freedom to multiply the function $f(z)$ with an arbitrary
constant without changing the result on the right hand side of Eq. 
(\ref{equ3}).  If we choose the constant carefully we can make the
contribution from the curved part of the contour vanish and we are only
left with an integration along the imaginary frequency axis:
\begin{equation}
 E\left( r \right) =  \frac{\hbar }{{4\pi }}\int\limits_{ -
 \infty }^\infty {d\omega } \ln f\left( {i\omega } \right), \\ 
\label{equ4}
\end{equation}
where the result was obtained from an integration by parts.  At finite
temperatures the force is \cite{Ser5}
\begin{equation}
f\left( r \right) =  - \frac{{d\mathfrak{F}}}
{{dr}},
\end{equation}
where Helmholtz' free energy is
\begin{equation}
\begin{gathered}
\mathfrak{F} = \sum\limits_i {\tfrac{1} {2}\varepsilon _i \left( r \right)
+ \tfrac{1} {\beta }\ln \left( {1 - e^{ - \beta \varepsilon _i \left( r
\right)} } \right)} \hfill \\ 
\quad = \sum\limits_i
{\tfrac{1} {\beta }\ln \left( {2\sinh \tfrac{1} {2}\beta \varepsilon _i }
\right)}. \hfill \\
\end{gathered} 
\end{equation}
We make the observation that we may use the generalized Argument Principle
but now with ${{\ln \left[ {2\sinh \left( {{{\beta \hbar z} \mathord{\left/
{\vphantom {{\beta \hbar z} 2}} \right.  \kern-\nulldelimiterspace} 2}}
\right)}
\right]} \mathord{\left/ {\vphantom {{\ln \left[ {2\sinh \left( {{{\beta
\hbar z} \mathord{\left/ {\vphantom {{\beta \hbar z} 2}} \right. 
\kern-\nulldelimiterspace} 2}} \right)} \right]} \beta }} \right. 
\kern-\nulldelimiterspace} \beta }$ instead of ${{\hbar z} \mathord{\left/
 {\vphantom {{\hbar z} 2}} \right.  \kern-\nulldelimiterspace} 2}$ for
 $\varphi {\rm{(z)}}$ in the integrand.  There is however one complication. 
 This new function has poles of its own in the complex frequency plane.  We
 have to choose our contour so that it includes all poles and zeros of the
 function $f(z)$ but excludes the poles of $\varphi {\rm{(z)}}$.  The poles of
 function $\varphi {\rm{(z)}}$ all are on the imaginary frequency axis.  We
 use the same contour as in Fig.  \ref{figu1}, but now let the straight
 part of the contour lie just to the right of, and infinitesimally close
 to, the imaginary axis.  We have
\begin{equation}
\begin{gathered}
  \mathfrak{F} = \frac{1}
{{2\pi i}}\int\limits_\infty ^{ - \infty } {d\left( {i\omega } \right)\frac{1}
{\beta }\ln \left( {2\sinh \tfrac{1}
{2}\beta \hbar i\omega } \right)\frac{d}
{{d\left( {i\omega } \right)}}\ln f\left( {i\omega } \right)}  \hfill \\
  \quad  = \frac{\hbar }
{{4\pi i}}\int\limits_{ - \infty }^{ + \infty } {d\omega \coth \left( {\tfrac{1}
{2}\beta \hbar i\omega } \right)\ln f\left( {i\omega } \right)}.  \hfill \\ 
\end{gathered} 
\end{equation}
The coth function has poles on the imaginary z-axis and they should not be
inside the contour.  The poles are at 
\begin{equation}
z_n  = i\frac{{2\pi n}}
{{\hbar \beta }}  ;  n = 0, \pm 1, \pm 2, \ldots ,
\end{equation}
and all residues are the same, equal to ${2 \mathord{\left/ {\vphantom {2
{\hbar \beta }}} \right.  \kern-\nulldelimiterspace} {\hbar \beta }}$.  We
integrate along the imaginary axis and deform the path along small
semicircles around each pole.  The integration path is illustrated in Fig. 
\ref{figu2}.  The integration along the axis results in zero since the
integrand is odd with respect to $\omega$.  The only surviving
contributions are the ones from the small semicircles.  The result is
\begin{equation}
\begin{gathered}
\mathfrak{F} = \frac{\hbar } {{4\pi i}}\sum\limits_{\omega _n }
{\frac{{i2\pi }} {2}\frac{2} {{\hbar \beta }}} \ln f\left( {i\omega _n }
\right) \hfill \\ \quad = \frac{1} {{2\beta }}\sum\limits_{\omega _n } {\ln
f\left( {i\omega _n } \right)} ; \omega _n = \frac{{2\pi n}} {{\hbar \beta
}} ; n = 0, \pm 1, \pm 2, \ldots \hfill \\
\end{gathered} 
\label{equ9}
\end{equation}
Since the summand is even in $n$ we can write this as
\begin{equation}
\mathfrak{F} = \frac{1}
{\beta }\sum\limits_{\omega _n } {^{'}\ln f\left( {i\omega _n } \right)} ;
\omega _n = \frac{{2\pi n}} {{\hbar \beta }} ; n = 0,1,2, \ldots ,
\end{equation}
where the prime on the summation sign indicates that the $n = 0$ term should
be multiplied by a factor of one half.  This factor of one half is because
there is only one term with $\left| n \right| = 0$ in the original summation but two for
all other integers.  When the temperature goes to zero the spacing between
the discrete frequencies goes to zero and the summation may be replaced by
an integration:
\begin{equation}
\begin{gathered}
\mathfrak{F} = \frac{1} {\beta }\sum\limits_{\omega _n }
{^{'}\ln f\left( {i\omega _n } \right)} \xrightarrow[{T \to
0}]{}\frac{{\hbar \beta }} {{2\pi }}\frac{1} {\beta }\int\limits_0^\infty
{d\omega } \ln f\left( {i\omega} \right) \hfill \\ \quad = \hbar
\int\limits_0^\infty {\frac{{d\omega }} {{2\pi }}} \ln f\left( {i\omega
} \right) = E, \hfill \\
\end{gathered}
\end{equation}
and we regain the contribution to the internal energy from the interactions,
the change in zero-point energy of the modes.
\begin{figure}
\includegraphics[width=4cm]{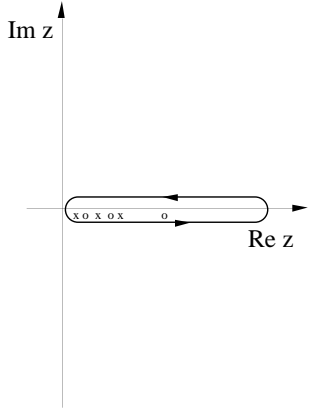}
\caption{Integration contour in the complex $z$-plane.  Crosses and circles
are poles and zeros, respectively, of the function $f(z)$.  The contour 
should be extended to include the whole positive part of the real axis. 
This contour is useful if one wants to find the contributions from 
different mode types}
\label{figu3}
\end{figure}

In all the three geometries we discuss in this work the interaction energy
per unit area, $V\left( d \right)$, can at zero temperature be written on
the form \cite{Ser}
\begin{equation}
 V\left( d \right) = \frac{\hbar }{\Omega }\sum\limits_{\bf{k}}
{\int\limits_0^\infty {\frac{{d\omega }}{{2\pi }}} } \ln \left[ {f\left(
{k,i\omega } \right)} \right],
\label{equ12}
\end{equation}
where $d$ is the distance between the objects, {\bf k} the two-dimensional
wave vector in the plane of the plate(s), $\Omega $
 the area of a plate, and
$f\left( {k,\omega } \right) = 0$ is the condition for an electromagnetic
normal mode in the particular geometry.  The integration is along the imaginary
frequency axis.  At finite temperature the integration is replaced by a
discrete summation over Matsubara frequencies,
\begin{equation}
V\left( d \right) = \frac{1}{{\beta \Omega }}\sum\limits_{\bf{k}}
{\sum\limits_{\omega _n }  }^{'} \ln \left[ {f\left( {k,i\omega _n } \right)}
\right];\;\omega _n = \frac{{2\pi n}}{{\hbar \beta }}.
\label{equ13}
\end{equation}
Alternatively one may integrate along the real frequency axis,
\begin{equation}
V\left( d \right) = \frac{{2\hbar }}{\Omega }\sum\limits_{\bf{k}}
{{\mathop{\rm Im}\nolimits} \int\limits_0^\infty {\frac{{d\omega }}{{2\pi
}}} } \left[ {n\left( \omega \right) + {1 \mathord{\left/ {\vphantom {1 2}}
\right.  \kern-\nulldelimiterspace} 2}} \right]\ln \left[ {f\left(
{k,\omega } \right)} \right],
\label{equ14}
\end{equation}
where $n\left( \omega \right) = \left[ {\exp \left( {\hbar \beta \omega }
\right) - 1} \right]^{ - 1} $ is the distribution function for massless
bosons.  This form can also be used at zero temperature; then the
distribution function vanishes.  These last results one arrives at by using
the contour in the complex frequency plane shown in Fig.  \ref{figu3}.
\begin{figure}
\includegraphics[width=8cm]{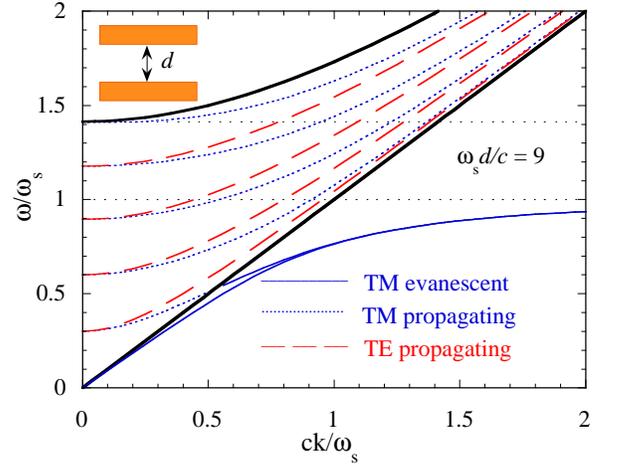}
\caption{Dispersion curves for the modes between two gold plates in 
absence of dissipation.  The frequencies are in units of $\omega _s $, the
surface plasmon frequency. The solid straight line is the light dispersion
curve in vacuum; the dashed (dotted) curves are TE (TM) propagating modes;
the thin solid curves are evanescent TM modes; the thick solid curve is the
lower boundary for transverse bulk modes in the plates.  From Ref. 
\cite{Ser2}}
\label{figu4}
\end{figure}
 The force per unit area, or pressure, is obtained as minus the derivative
 with respect to distance, $F\left( d \right) = - {{{\rm{d}}V\left( d
 \right)} \mathord{\left/ {\vphantom {{{\rm{d}}V\left( d \right)}
 {{\rm{d}}d}}} \right.  \kern-\nulldelimiterspace} {{\rm{d}}d}}$.

In all three geometries there are two
groups of normal mode, transverse magnetic (TM) and transverse electric
(TE), each with a different mode condition function.  The interaction
potential is a sum of two terms, $ V\left( d \right) = V^{TM} \left( d
\right) + V^{TE} \left( d \right) $.  In the two plate geometries the mode condition
functions are
\begin{equation}
  f^{\scriptstyle TM, \hfill \atop \scriptstyle TE \hfill} \left( {k,\omega
  } \right) = 1 - e^{ - 2\gamma _0 \left( {k,\omega } \right)d}
  r_{01}^{\scriptstyle TM, \hfill \atop \scriptstyle TE \hfill} \left(
  {k,\omega } \right)r_{02}^{\scriptstyle TM, \hfill \atop \scriptstyle TE
  \hfill} \left( {k,\omega } \right),
\end{equation}
where the Fresnel amplitude reflection coefficients at an interface between 
medium {\it i} and {\it j} are
\begin{equation}
r_{ij}^{TM} \left( {k,\omega } \right) = \frac{{\varepsilon _j \left(
\omega \right)\gamma _i \left( {k,\omega } \right) - \varepsilon _i \left(
\omega \right)\gamma _j \left( {k,\omega } \right)}}{{\varepsilon _j \left(
\omega \right)\gamma _i \left( {k,\omega } \right) + \varepsilon _i \left(
\omega \right)\gamma _j \left( {k,\omega } \right)}},
\end{equation}
for TM modes (p-polarized waves) and
\begin{equation}
r_{ij}^{TE} \left( {k,\omega } \right) = \frac{{\gamma _i \left( {k,\omega
} \right) - \gamma _j \left( {k,\omega } \right)}}{{\gamma _i \left(
{k,\omega } \right) + \gamma _j \left( {k,\omega } \right)}},
\end{equation}
for TE modes (s-polarized waves), respectively.  We have let the objects be of medium 1
and 2 and let the surrounding vacuum be denoted by medium 0.  The gamma
functions are
\begin{equation}
\gamma _j \left( {k,\omega } \right) = \sqrt {k^2 - \varepsilon _j \left(
\omega \right)\left( {{\omega \mathord{\left/
 {\vphantom {\omega  c}} \right.
 \kern-\nulldelimiterspace} c}} \right)^2 } ;\;j = 0,1,2.
\end{equation}

Now we have all material we need for now. The interaction potential for the 
atom wall geometry can also be derived from these relations using a limit 
procedure. We make that derivation in Sec. \ref{G3}.

\section{\label{G1}Two parallel metal plates}

The dispersion curves for the electromagnetic normal modes for two gold
plates \cite{Ser2} is shown in Fig.  \ref{figu4} .  This figure is valid in
neglect of dissipation in the plate materials.  The modes are propagating
(evanescent) above and to the left (below and to the right) of the light
dispersion curve.  The light dispersion curve is the straight diagonal line
in the figure; it has slope unity with the chosen scaling of the axes. 
Note that there are no TE evanescent modes.  When the system is allowed to
have dissipation there are modes everywhere.  Each original mode is
replaced by a continuum of modes
\cite{Ser3}.  Evanescent TE modes appear and the continuum extends all the
way down to the momentum axis.  These modes are the cause of all the
problems with the thermal Casimir force in this geometry. 
\begin{figure}
\includegraphics[width=8cm]{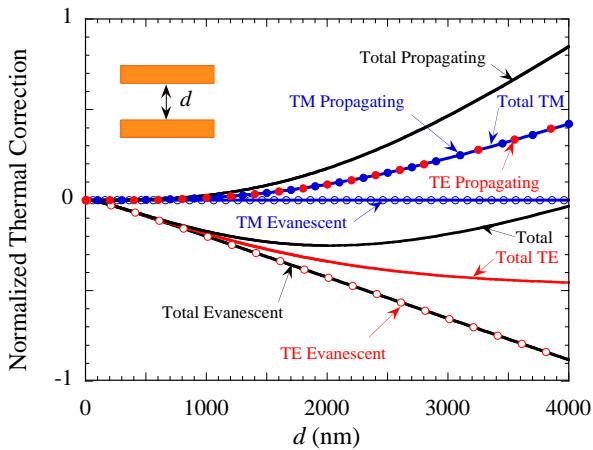}
\caption{The thermal contribution to the Casimir energy from the four mode 
types between two gold plates. The results are also lumped together in 
various ways. All results are divided by the zero temperature Casimir 
energy for perfectly reflecting plates. Similar results have been obtained 
earlier for silver plates \cite{Bos}}
\label{figu5}
\end{figure}

In Fig.  \ref{figu5} we show the thermal correction to the Casimir energy. 
The contribution from all four mode types are given separately. These
curves have been obtained from Eq.  (\ref{equ14}) and the Drude dielectric
function suited for gold has been used. We note that the TE evanescent waves
give a negative contribution.  For larger separations the total TE
contribution saturates when it has completely eliminated the zero
temperature TE contribution.

\begin{figure}
\includegraphics[width=8cm]{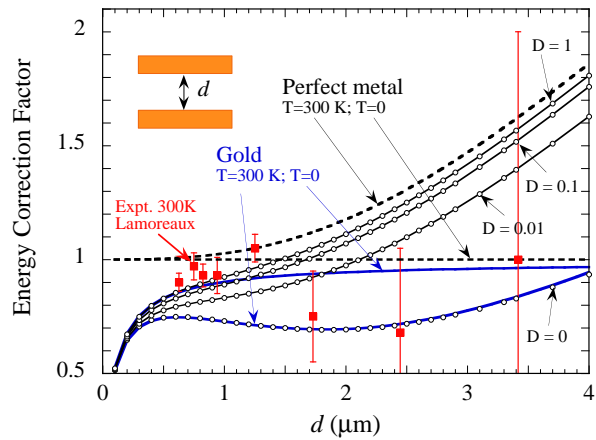}
\caption{Energy correction factor for two gold plates. The filled squares 
with error bars are the Lamoreaux' experimental \cite{Lamo} values from the
torsion pendulum experiment.  The dashed curves are the perfect metal
results.  The thick solid curves are the results for real gold plates using Eq. 
(\ref{equ12}) for zero temperature and Eq. (\ref{equ13}) for room 
temperature. The dielectric properties for gold was obtained from tabulated
experimental optical data. The curves with circles are the results from our 
model calculations with different saturation parameters.}
\label{figu6}
\end{figure}
 In Fig.  \ref{figu6} we show the energy correction factor for the
 two-gold-plate geometry.  The energy correction factor is the Casimir
 energy per unit area divided by the zero temperature Casimir energy per
 unit area for two perfectly reflecting metal plates, ${{ - \hbar c\pi ^2 }
 \mathord{\left/ {\vphantom {{ - \hbar c\pi ^2 } {\left( {720d^3 }
 \right)}}} \right.  \kern-\nulldelimiterspace} {\left( {720d^3 }
 \right)}}$.  The experimental results \cite{Lamo} obtained by Lamoreaux
 are shown as squares with error bars.  The larger the separation the
 bigger the errors.  There is a cluster of experimental points near $1\mu
 m$ separation.  Here the deviation between theory and experiment is over
 $20\% $.  Theory and experiment are in clear disagreement.  The proposed
 prescription has been to neglect the dissipation in the intraband part of
 the dielectric function but keep it in the interband part
\cite{MosGey}. 
 
Let us now explain our view of what goes wrong in the theory of the thermal
Casimir effect in presence of dissipation.  The traditional theory relies
fully on the concept of electromagnetic normal modes.  These are assumed to
be independent massless bosons.  The possibility to excite one of these
modes is assumed to be completely independent of how many modes are already
excited.  An excitation of a mode involves excitations of the charged
particles in the system, electrons in the geometries studied here.  These
are the sources of the fields.  Now, the electrons are fermions and there
is at most one electron in each particle state.  An electron that is
excited at one instant of time cannot immediately be excited again -- the
state is empty.  The more modes that are excited the more difficult it is
to excite new modes --- there are saturation effects.  It is a matter of
logistics.  In the theoretical treatment this is not taken care of.  In
most cases this fact will not cause any problems, but sometimes it could. 
We think that the thermal Casimir effect is one such case.  When
dissipation is included each mode is replaced by a continuum of an infinite
number (for an infinite system) of new modes.  The distribution function
diverges towards zero frequency and the saturation effects should appear
here.  This is very difficult to treat in a strict way.  We use an
approximation which is very easy to implement.  We shift the distribution
function in Eq.  (\ref{equ14}) downwards in frequency, so that it never
reaches the point of divergence, by adding a damping parameter, $D$,
\begin{equation}
 \tilde n\left( \omega \right) = \left[ {\exp \left( {\hbar \beta \omega +
 D} \right) - 1} \right]^{ - 1}.
\label{equ19}
\end{equation}
 The discrete frequency summation in Eq.  (\ref{equ13}) is the result of the
 poles of the distribution function that all fall on the imaginary axis,
 see Ref.  \cite{Ser}.  Our new distribution function has its poles shifted
 away from the axis the distance ${D \mathord{\left/ {\vphantom {D {\hbar
 \beta }}} \right.  \kern-\nulldelimiterspace} {\hbar \beta }}$ into the
 left half plane. The new form of the interaction potential is
\begin{equation}
V\left( d \right) = \frac{1}{{\beta \Omega }}\sum\limits_{\bf{k}}
{\sum\limits_{\omega _n } }^{' } \frac{1}{\pi }\int\limits_{ - \infty }^\infty
{d \omega ' \frac{{\left( {{D \mathord{\left/ {\vphantom {D \beta }} \right. 
\kern-\nulldelimiterspace} \beta }} \right)\ln \left[ {f\left( {k,i\omega
'} \right)} \right]}}{{\left( {\omega ' - \omega _n } \right)^2 + \left(
{{D \mathord{\left/ {\vphantom {D \beta }} \right. 
\kern-\nulldelimiterspace} \beta }} \right)^2 }}}. 
\label{equ20}
\end{equation}
Each term in the summation is replaced by an integral. For small $D$ 
values it is enough to replace only the zero frequency term. 
\begin{figure}
\includegraphics[width=8cm]{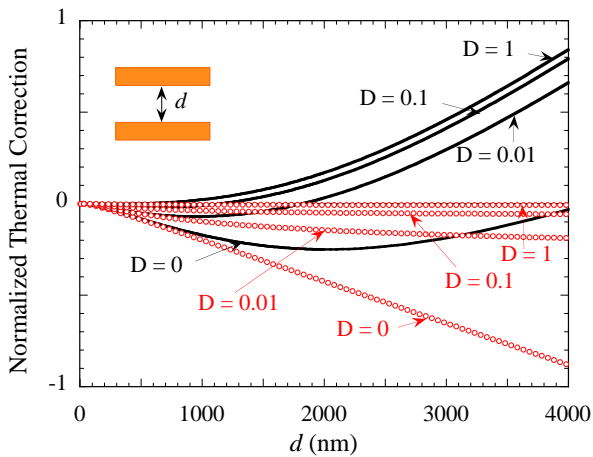}
\caption{The thermal contribution to the Casimir energy from the TE 
evanescent modes between two gold plates for different damping parameters,
circles.  The corresponding total corrections are shown as solid curves. 
Only the TE evanescent modes have been modified by the saturation.}
\label{figu7}
\end{figure}

The circles in Fig. (\ref{figu6}) are 
the results with damping parameters 0, 0.01, 0.1, and 1.0, 
respectively, counting from below. We have used Eq. (\ref{equ14}) with the 
modified distribution function to get the thermal correction.
A word of caution is in place. This approach changes the population of 
modes at all energies, not just at the low frequency limit where 
saturation effects are predicted to appear. To limit the damage of this we have 
just taken into account the change in the contribution from TE evanescent 
waves. These contributions are the dominating ones at low energies. In 
Fig. \ref{figu7} we show the thermal correction from the TE evanescent 
waves, including the damping, as circles and the total correction as solid 
curves.

\begin{figure}
\includegraphics[width=8cm]{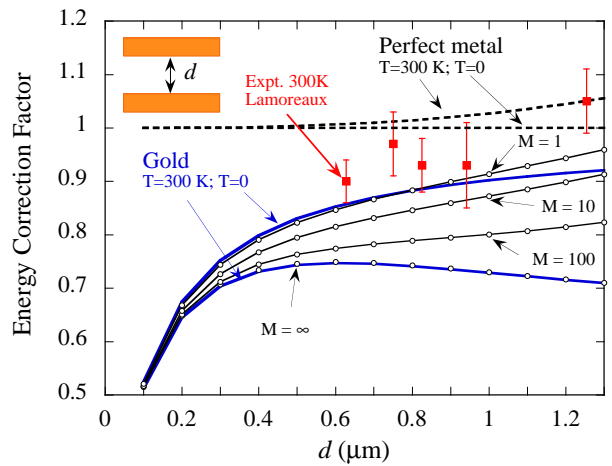}
\caption{Energy correction factor for two gold plates. The filled squares 
with error bars are the Lamoreaux' experimental \cite{Lamo} values from the
torsion pendulum experiment.  The dashed curves are the perfect metal
results.  The thick solid curves are the results for real gold plates using Eq. 
(\ref{equ12}) for zero temperature and Eq. (\ref{equ13}) for room 
temperature. The dielectric properties for gold was obtained from tabulated
experimental optical data. The curves with circles are the results from our 
model calculations with different saturation strengths.  Here the 
distribution function for all mode types is limited to the value M. The 
Drude dielectric function was used in the calculation of the thermal correction}
\label{figu8}
\end{figure}

An alternative way to treat the saturation is to simply introduce a cutoff
in the distribution function.  In Fig.  \ref{figu8} we show the results for
the energy correction factor from using Eq.  (\ref{equ14}) with a cutoff. 
The parameter M is the maximum allowed value of the distribution function. 
We have here limited the separation range to focus on the interesting
cluster of experimental points near $1\mu m$ separation.  This cutoff
method forces us to stay on the real frequency axis where the results are
more sensitive to variations in the dielectric function and the numerical
calculation more difficult to perform.

\begin{figure}
\includegraphics[width=8cm]{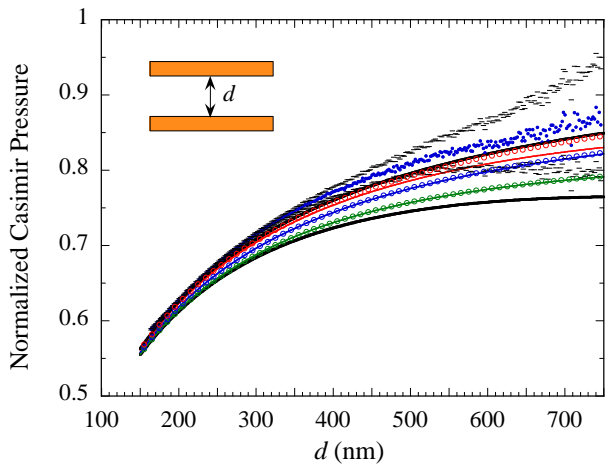}
\caption{Casimir pressure between two gold plates. The experimental result 
from Ref.  \cite{Decca} is shown as dots and the endpoints of the error
bars are indicated by horizontal bars; the upper (lower) thick solid curve
is the traditional theoretical zero (room) temperature result; the circles
are the present results, obtained from Eq.  (\ref{equ14}) with the
distribution function for TE evanescent waves modified according to Eq. 
(\ref{equ19}), with damping parameters 0.01, 0.1, and 1.0, respectively,
counting from below; the corresponding results obtained by shifting the
zero frequency pole of the cosh function into the left of the complex
frequency plane are shown as thin solid curves.}
\label{figu9}
\end{figure}
The experimental result \cite{Decca} for the normalized Casi-mir pressure
at 295 K is shown as dots in Fig.  \ref{figu9}.  The bars are the endpoints
of the experimental error bars.  The upper (lower) thick solid curve is the
theoretical result for zero temperature (295 K) calculated with Eqs. 
(\ref{equ12}) and (\ref{equ13}), respectively.  The dielectric function on
the imaginary frequency axis was derived from experimental tabulated
optical data for gold.  We note that the zero temperature result agrees
much better with the experimental result.  The large negative thermal
correction comes entirely from the TE evanescent modes \cite{TorLam}.  All
curves are normalized with the zero temperature Casimir pressure between
two perfectly reflecting metal plates, ${{\hbar c\pi ^2 } \mathord{\left/
{\vphantom {{\hbar c\pi ^2 } {\left( {240z^4 } \right)}}} \right. 
\kern-\nulldelimiterspace} {\left( {240z^4 }
\right)}}$.  We have neglected surface roughness effects. The circles are 
the results for different damping parameters from using Eq.  ({\ref{equ14})
with the modified distribution function of Eq.  ({\ref{equ19}) in the
contribution from the TE evanescent waves.  To each set of circles
corresponds a thin solid curve.  This curve is the result of using Eq. 
(\ref{equ13}), where just the zero frequency pole of the cosh function has
been moved into the left half of the complex frequency plane and the
corresponding term in the summation has been modified according to Eq. 
(\ref{equ20}).  We note that for the two lowest set of curves with small
damping the two results agree.  For very high damping there are deviations. 
These deviations have two reasons: One is that in the thin solid curves all
mode types are affected by the damping; the second is that for strong
enough damping more terms in the summation, leading to the sets of circles,
should be modified.

\section{\label{G2}Two parallel plates, one metallic and one semiconducting}
\begin{figure}
\includegraphics[width=8cm]{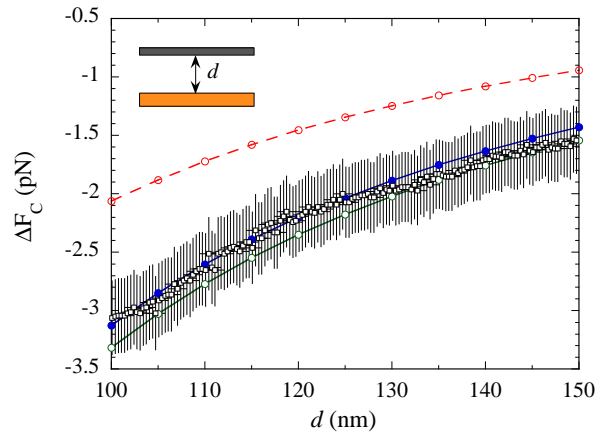}
\caption{The change in Casimir force, at 300 K, between a gold sphere and a silicon 
membrane with and without laser irradiation.  The open squares with error
bars are the experimental \cite{Chen} result.  The dashed curve with
open circles is the theoretical result without saturation effects.  The solid curve
with filled (open) circles is our present result with $D$ equal to 0.01
(0.1). }
\label{figu10}
\end{figure}

The third experiment we consider here is the measurement of the Casimir
force between a gold plate and a laser irradiated semiconductor membrane as
performed by Chen et al.  \cite{Chen}.  They measured the change in force
with the laser irradiation compared to without any irradiation.  The idea
is to find out how the force varies with carrier concentration in the
semiconducting membrane.  The results are shown in Fig.  \ref{figu10}.  The
open squares with error bars are the experimental result.  The dashed curve
with open circles is the theoretical result for 300 K. The deviations are
clear.  In this geometry it is not enough to neglect dissipation to get
agreement with experiment; the theoretical results with and without
dissipation are very similar.  Besides, it is now the TM modes that cause
the problems.  In the non-irradiated semiconductor material there is always
some thermally excited carriers present.  One postulated that for the
non-irradiated semiconductor one should completely omit the contribution to
the dielectric function from the thermally excited carriers.  That brought
theory and experiment into agreement, see Fig.  10 in Ref.
\cite{Chen}.  In the present Fig.  \ref{figu10} the solid curve with filled
(open) circles is our saturation based result with $D$ equal to 0.01 (0.1). 
We find that both these curves agree with the experiment within the
experimental uncertainty.  Here we have used Eq.  (\ref{equ13}) and just
modified the zero frequency contribution according to Eq.  (\ref{equ20}). 
We have derived the dielectric function for the gold plate from
experimental tabulated optical data.  For the silicon with and without
laser irradiation we have used the dielectric function given in Ref. 
\cite{Chen}.

\section{\label{G3}Atom wall geometry}

The Casimir force between an atom and a wall can be obtained from the
results of the two plate geometry.  One takes the limit when the thickness
of one of the plates goes to zero and at the same time lets the material of
the thin plate be diluted.  We start with a half space of medium 1 and a
layer of medium 2 emersed in vacuum (medium 0) separated by the distance
$d$. The function in the condition for electromagnetic normal modes is 
then given by
\begin{equation}
f = 1 + e^{ - 2\gamma _0 d} \frac{{r_{10} r_{02} + e^{ - 2\gamma _2 d_2 }
r_{10} r_{20} }} {{1 + e^{ - 2\gamma _2 d_2 } r_{02} r_{20} }},
\end{equation}
where we have suppressed the arguments $k$ and $\omega$.  We have
furthermore divided the original function with the one corresponding to infinite
separation.  This corresponds to choosing our reference system to be the one
at infinite separation. There will be one function like this for TM modes 
and one for TE modes. 
Using that the thickness of layer 2 is small and that the material in this 
layer is diluted with a small reflection coefficient as a result leads to

\begin{equation}
\begin{gathered}
  f = 1 + e^{ - 2\gamma _0 d} r_{10} r_{02} \frac{{1 - e^{ - 2\gamma _2 d_2
  } }} {{1 - e^{ - 2\gamma _2 d_2 } \left( {r_{02} } \right)^2 }} \hfill \\
  \quad \approx 1 + e^{ - 2\gamma _0 d} r_{10} r_{02} 2\gamma _2 d_2 . 
  \hfill \\
\end{gathered} 
\end{equation}
We now let the dielectric function in the layer be $\varepsilon _2 = 1 +
4\pi n\alpha \left( \omega \right) \approx 1 + {{4\pi \alpha \left( \omega
\right)} \mathord{\left/ {\vphantom {{4\pi \alpha \left( \omega \right)}
{\Omega d_2 }}} \right.  \kern-\nulldelimiterspace} {\Omega d_2 }}$.  We
have assumed that there is only one atom in the layer, with polarizability
${\alpha \left( \omega \right)}$.  At this point we need to specify the
mode type.  We start with TE modes:

\begin{equation}
f^{TE} \approx 1 + e^{ - 2\gamma _0 d} \frac{{\gamma _1 - \gamma _0 }}
{{\gamma _1 + \gamma _0 }}\frac{{\gamma _0 - \gamma _2 }} {{\gamma _0 +
\gamma _2 }}2d_2 \gamma _2.
\end{equation}
We furthermore have
\begin{equation}
\begin{gathered}
 \gamma _2 = \sqrt {k^2 - \left( {{\omega \mathord{\left/ {\vphantom
 {\omega c}} \right.  \kern-\nulldelimiterspace} c}} \right)^2 - {{4\pi
 \alpha \left( \omega \right)\left( {{\omega \mathord{\left/ {\vphantom
 {\omega c}} \right.  \kern-\nulldelimiterspace} c}} \right)^2 }
 \mathord{\left/ {\vphantom {{4\pi \alpha \left( \omega \right)\left(
 {{\omega \mathord{\left/ {\vphantom {\omega c}} \right. 
 \kern-\nulldelimiterspace} c}} \right)^2 } {\Omega d_2 }}} \right. 
 \kern-\nulldelimiterspace} {\Omega d_2 }}} \hfill \\ \quad \approx \gamma
 _0 \left( {k,\omega } \right) - {{2\pi \alpha \left( \omega \right)\left(
 {{\omega \mathord{\left/ {\vphantom {\omega c}} \right. 
 \kern-\nulldelimiterspace} c}} \right)^2 } \mathord{\left/ {\vphantom
 {{2\pi \alpha \left( \omega \right)\left( {{\omega \mathord{\left/
 {\vphantom {\omega c}} \right.  \kern-\nulldelimiterspace} c}} \right)^2 }
 {\Omega d_2 \gamma _0 \left( {k,\omega } \right)}}} \right. 
 \kern-\nulldelimiterspace} {\Omega d_2 \gamma _0 \left( {k,\omega }
 \right)}}.  \hfill \\
\end{gathered} 
\end{equation}
Thus, the result is
\begin{equation}
f^{TE} \approx 1 + \frac{1} {\Omega }e^{ - 2\gamma _0 d} \frac{{\gamma _1 -
\gamma _0 }} {{\gamma _1 + \gamma _0 }}\frac{{2\pi \alpha \left( \omega
\right)\left( {{\omega \mathord{\left/ {\vphantom {\omega c}} \right. 
\kern-\nulldelimiterspace} c}} \right)^2 }} {{\gamma _0 }}.
\end{equation}
For the TM modes we have
\begin{equation}
f^{TM} \approx 1 + e^{ - 2\gamma _0 d} \frac{{\gamma _1 - \varepsilon _1
\gamma _0 }} {{\gamma _1 + \varepsilon _1 \gamma _0 }}\frac{{\varepsilon _2
\gamma _0 - \gamma _2 }} {{\varepsilon _2 \gamma _0 + \gamma _2 }}2d_2
\gamma _2,
\end{equation}
and in the limit
\begin{equation}
f^{TM} \approx 1 + \frac{1} {\Omega }e^{ - 2\gamma _0 d} \frac{{\gamma _1 -
\varepsilon _1 \gamma _0 }} {{\gamma _1 + \varepsilon _1 \gamma _0
}}\frac{{2\pi \alpha \left( \omega \right)\left[ {2k^2 - \left( {{\omega
\mathord{\left/ {\vphantom {\omega c}} \right.  \kern-\nulldelimiterspace}
c}} \right)^2 } \right]}} {{\gamma _0 }}.
\end{equation}
In this geometry we want the interaction energy, not the interaction energy
per unit area, so we multiply with the area of the plate and find
\begin{equation}
\begin{gathered}
V\left( d \right) = \frac{\hbar } {\Omega }\sum\limits_{\mathbf{k}}
{\int\limits_0^\infty {\frac{{d\omega }} {{2\pi }}} \left[ {\Omega \ln
f^{TE} \left( {{\mathbf{k}},i\omega } \right)} \right.} \hfill \\ \quad
\quad \quad + \left.  {\Omega \ln f^{TM} \left( {{\mathbf{k}},i\omega }
\right)} \right], \hfill \\
\end{gathered}
\end{equation}
for zero temperature and
\begin{equation}
\begin{gathered}
V\left( d \right) = \frac{\hbar } {{\beta \Omega }}\sum\limits_{\mathbf{k}}
{{\sum\limits_{\omega _n }}^{'} {\left[ {\Omega \ln f^{TE} \left(
{k,i\omega _n } \right)} \right.} } \hfill \\ \quad \quad \quad \quad
+ \left.  {\Omega \ln f^{TM} \left( {k,i\omega _n } \right)}
\right], \hfill \\
\end{gathered} 
\end{equation}
for finite temperature, where
\begin{equation}
\Omega \ln f^{TE} \left( {{\mathbf{k}},\omega } \right) \approx \alpha e^{
- 2\gamma _0 d} \frac{{\gamma _1 - \gamma _0 }}
{{\gamma _1  + \gamma _0 }}\frac{{2\pi \left( {{\omega  \mathord{\left/
 {\vphantom {\omega  c}} \right.
 \kern-\nulldelimiterspace} c}} \right)^2 }}
{{\gamma _0 }},
\end{equation}
and
\begin{equation}
\Omega \ln f^{TM} \left( {{\mathbf{k}},\omega } \right) \approx \alpha e^{
- 2\gamma _0 d} \frac{{\gamma _1 - \varepsilon _1 \gamma _0 }}
{{\gamma _1 + \varepsilon _1 \gamma _0 }}\frac{{2\pi \left[ {2k^2 - \left(
{{\omega \mathord{\left/ {\vphantom {\omega c}} \right. 
\kern-\nulldelimiterspace} c}} \right)^2 } \right]}} {{\gamma _0 }},
\end{equation}
respectively.  We note that the zero frequency contribution vanishes for the
TE modes, but not for the TM modes.  The force between the atom and the
wall is
\begin{equation}
\begin{gathered}
  F\left( d \right) = - \frac{{4\pi \hbar }} {{\beta \Omega
  }}\sum\limits_{\mathbf{k}} {\sum\limits_{\omega _n } }^{'} \left[ {\alpha
  e^{ - 2\gamma _0 d} \frac{{\gamma _1 - \gamma _0 }} {{\gamma _1 + \gamma
  _0 }}\left( {{{\omega _n } \mathord{\left/ {\vphantom {{\omega _n } c}}
  \right.  \kern-\nulldelimiterspace} c}} \right)^2 } \right.  \hfill \\
  \quad \quad \quad \quad \quad - \left.  {\alpha e^{ - 2\gamma _0 d}
  \frac{{\gamma _1 - \varepsilon _1 \gamma _0 }} {{\gamma _1 + \varepsilon
  _1 \gamma _0 }}\left[ {2k^2 + \left( {{{\omega _n } \mathord{\left/
  {\vphantom {{\omega _n } c}} \right.  \kern-\nulldelimiterspace} c}}
  \right)^2 } \right]} \right], \hfill \\
\end{gathered} 
\label{equ32}
\end{equation}
where the first term is the TE contribution and the second the TM 
contribution. The suppressed arguments of the functions in the integrand 
are $\left( {k,i\omega _n } \right)$.

 In the experiment by Obrecht et al.  \cite{Obrecht} one studied indirectly
 the force between a rubidium atom and a dielectric substrate.  This was
 done by measuring the collective oscillation frequency of the mechanical
 dipole mode of a BEC near enough to a dielectric substrate for the force
 to measurably distort the trapping potential.  In that work one studied
 both the equilibrium case when the substrate was of the same temperature
 as the surrounding environment and the non-equilibrium case when the
 substrate was of a different temperature.  We concentrate on the
 equilibrium case when both the surroundings and the wall have the same
 temperature, 310 K. The fractional change in trap frequency is defined
 as $\gamma _x \equiv {{\left( {\omega _0 - \omega _x } \right)}
 \mathord{\left/ {\vphantom {{\left( {\omega _0 - \omega _x } \right)}
 {\omega _0 }}} \right.  \kern-\nulldelimiterspace} {\omega _0 }}$ in terms
 of the unperturbed trap frequency, ${\omega _0 }$, and ${\omega _x }$, the
 trap frequency perturbed by the force. This may be obtained as
\begin{equation}
\gamma _x  = \frac{1}
{{ma\omega _0^2 }}\left| {\Phi _e \left( d \right)} \right|,
\end{equation}
where the expression for ${\Phi _e \left( d \right)}$ is identical to Eq. 
(\ref{equ32}) except that the integrand or summand has an extra factor $I_1
\left( {2\gamma _0 a} \right)g\left( {2\gamma _0 R_x } \right)$. In the 
experiment $a = 2.50\mu m$ is the amplitude of oscillation, $R_x = 2.69\mu
m$ is the Thomas-Fermi radius in the $x$-direction and $m = 1.443 \times
10^{ - 25} kg$ is the mass of the rubidium atom.  The function $I_1 \left(
z \right)$ is the Bessel function and $g\left( z \right) \equiv {{15\left[
{\left( {3 + z^2 } \right)\sinh z - 3z\cosh z} \right]} \mathord{\left/
{\vphantom {{15\left[ {\left( {3 + z^2 } \right)\sinh z - 3z\cosh z}
\right]} {z^5 }}} \right.  \kern-\nulldelimiterspace} {z^5 }}$.  These
results were obtained in Ref.
\cite{KliMos} after an analytical averaging procedure had been performed.
\begin{figure}
\includegraphics[width=8cm]{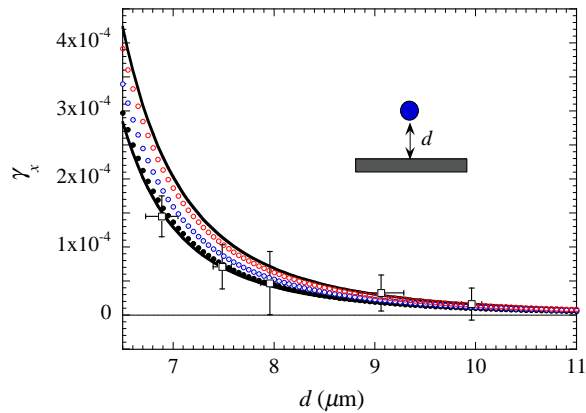}
\caption{Fractional change in trap frequency for a Rb atom near a silica 
wall versus separation in thermal equilibrium.  The open squares are the
experimental result \cite{Obrecht}.  The upper (lower) curve is the theoretical result
including (neglecting) the conductivity from the few thermal carriers in the
silica wall.  The circles are our present results for the $D$ values $10^{
- 10}$, $10^{ - 11}$, and $10^{ - 12}$, respectively, counted from below.}
\label{figu11}
\end{figure}

In Fig.  \ref{figu11} the experimental result \cite{Obrecht} is shown as
open squares with error bars. The upper (lower) curve is the theoretical result,
without saturation, including (neglecting) the conductivity from the few
thermal carriers in the silica wall.  We see that also here the neglect of
the contribution, to the dielectric function of the silica wall, from the
very few thermally excited carriers brings the theoretical result into
agreement with experiment.  This neglect is the postulated remedy in Ref.
\cite{KliMos}.  In this geometry, just as in gold-plate silicon-wafer 
geometry, the TM modes cause the problems and it is not enough to neglect
dissipation to get good agreement between theory and experiment.  To
include saturation effects we have just modified the zero frequency
contribution in analogy with Eq.  (\ref{equ20}).  We note that in this
experiment it is enough to have a damping parameter as small as $10^{ -
10}$ to bring the theoretical result into agreement with experiment.  We
have assumed that the thermally excited carriers in the wall material have
the conductivity $100\,s^{ - 1}$ ($ \sim 10^{ - 10} {\rm{ohm}}^{ - 1}
{\rm{cm}}^{ - 1}$), which is the upper limit of the range given in Ref. 
\cite{KliMos}.  Using smaller values leads to even weaker demands on the
damping parameter.

\section{\label{summary}summary and discussions}

In summary we have proposed that saturation effects are responsible for the
discrepancy between theory and experiment in several quite different
Casimir geometries.   It is quite difficult or even impossible to derive the saturation 
effects from first principles. We have treated saturation within a very simple
calculation model and demonstrated that the problems may go away in all
cases. The basic assumption is that the electromagnetic normal modes can 
not be fully treated as independent massless bosons. There are limitations 
on the number of excited modes and this limitation shows up at low 
frequencies where the distribution function for massless bosons diverges in 
the case of non-zero temperature. We encourage experimentalists to design 
experiments where the occupation of these modes is directly investigated.

\begin{acknowledgments}
We are grateful to R.S. Decca, G.L. Klimchitskaya, and U. Mohideen for
providing us with experimental data.  The research was sponsored by the
VR-contract No:70529001 and support from the VR Linn\'{e} Centre LiLi-NFM
and from CTS is gratefully acknowledged.
\end{acknowledgments}

\end{document}